\documentclass[a4paper,10pt]{article}
\usepackage{multirow}
\usepackage{array}
\usepackage{cancel}
\usepackage{amsmath}
\usepackage{amsthm}
\usepackage{amssymb}
\usepackage{hyperref}
\usepackage[usenames,dvipsnames]{color}
\usepackage[all]{xy}
\usepackage{graphicx}

\begin{document}

\title{Probing the flavor of the top quark decay}

\author{P.~Silva, M.~Gallinaro
\footnote{Laborat\'{o}rio de Instrumenta\c{c}\~{a}o e F\'{i}sica Experimental de Part\'{i}culas (LIP) - Av. Elias Garcia 14 - 1$^0$ 1000-149 Lisboa Portugal.}\\ \\
{\small \it Published Il Nuovo Cimento Vol. 125 B (August 2010).}\\
{\small \it The original method presented in this manuscript is}\\
{\small \it partially described in P. Silva's PhD thesis (ref. \cite{Psilva:2009th}.)}\\ \\
%}
}

\maketitle

\begin{abstract}
The top quark sector is almost decoupled from lighter quark generations due to the fact that $V_{tb}\approx$~1.
The current experimental measurements of $V_{tb}$ are compatible with the Standard Model expectations but are still dominated by experimental uncertainties.
In this manuscript, a revision of the experimental methods used to measure $V_{tb}$
is given, and a simple method to probe heavy flavor content fraction of top quark events, $R=B(t\rightarrow Wb)/B(t\rightarrow Wq)$, 
is presented and discussed.
Prospects for the measurements at the Large Hadron Collider based on generator level simulations are outlined.
\end{abstract}

\section{Introduction}
\label{sec:introduction}

One of the yet unveiled mysteries in nature is the mechanism that gives mass to 
quarks, leptons and gauge bosons breaking the electroweak symmetry.
The top quark, the heaviest elementary fermion observed, constitutes
{\it per s\'e} a puzzle in this context not only due to its large mass (close to a Yb atom)
but also due to the fact that it is practically decoupled from the
lighter quark generations decaying almost solely into the $t\rightarrow Wb$ channel.
The study of top quark events and its properties is therefore
potentially an interesting topic and one of the last barriers
to cross in the search for physics beyond the Standard Model (SM) of interactions.

In the context of the SM and at hadron colliders, such as the Tevatron or the Large Hadron Collider (LHC), 
top quarks are produced through the interaction of quarks and gluons, either in pairs or as single Top production (through the $s-$, $t-$, $tW-$channels).
While $t\bar{t}$ pairs are mostly created from pure QCD processes,
single top quark production is more often produced from electroweak processes and its production cross section 
is directly proportional to the (square of the) CKM matrix elements related to the top quark, i.e. $V_{tq}$.
Therefore, the measurement of single Top cross section in the different channels provides a direct measurement of the CKM matrix elements.
However, results from single Top production are still constrained mainly due to the limited number of events 
(at the Tevatron, the single Top cross section is $\sim$2.8~pb, while $t\bar{t}$ pair production has a cross section of $\sim$7.5~pb) 
and to the larger relative background contribution.
In practice, single top quark discovery has only been reported recently.
The value of $V_{tb}$ is measured with an 8\% relative uncertainty, assuming that $V_{ts}$, $V_{td} \ll V_{tb}$~\cite{singleTop}~\cite{Group:2009qk}. 
A more accurate determination of the separate contributions of the different channels is expected to be feasible
at the LHC, 
and the sensitivity to the $tW-$channel will be in the reach of the experiments for the first time.
A direct determination of the $t-$channel cross section is expected to yield a measurement 
of $V_{tb}$ with an overall uncertainty of 5\% after 10~fb$^{-1}$~\cite{Alwall:2006bx}.
Top quark pairs may provide an indirect measurement of $V_{tb}$, through the flavor determination of the Top quark decay products.
In particular, 
the ratio of branching fractions of the the top quark

\begin{equation}
R=\frac{B(t\rightarrow Wb)}{\displaystyle{\sum_q}B(t\rightarrow Wq)}
\label{eq:r} 
\end{equation}

can be measured in the data and, by assuming the unitarity of the CKM matrix elements, one can infer $|V_{tb}|^2$ from 
Eq.~\ref{eq:r} and derive $V_{tb}=\sqrt{R}$.
This indirect measurement has been explored at the Tevatron~\cite{Abazov:2008yn,Acosta:2005hr}.
The measurement using $t\bar{t}$ pair events profits from the larger production cross section
($\sim$7.5~pb at the Tevatron and $\sim$908~pb for the LHC at $\sqrt{s}$=14~TeV) and from the distinctive $t\bar{t}$ experimental signature 
with a much reduced relative background contamination.

Results always suffer however from the systematic uncertainties related to: i) the imperfect knowledge of the efficiency
of the algorithms used to identify the flavor of jets (``b-tagging''), and to ii) the background modeling.
In the current Tevatron results, $R$ is determined with a relative uncertainty of $\sim$9\%~\cite{Abazov:2008yn}.
Notice that $R\approx 1$ is a direct indication of the strong hierarchy between the top quark 
sector and the lighter quark generations. 
However, $R\approx 1$ can still be compatible with $V_{tb}<1$ if the structure of the CKM
matrix is different from the one assumed in the SM (e.g. if a fourth family of quarks exists
or if the top quark sector couples to a massive vector like quark $t'$).
The current measurements of $R$ and $V_{tb}$ have not yet reached the required precision to discard such hypotheses.
Therefore the measurement of both $R$ and the deviations in the heavy flavor content of $t\bar{t}$
events are particularly important to search for evidence of New Physics at the LHC.
A brief summary of the SM experimental signature of the top quark events 
and examples of how the $t\bar{t}$ sample can be contaminated by non-SM processes are given in Sec.~\ref{subsec:topexperimentalsignature}.
The discussion of the method and how the measurement of $R$ and of the cross section can be performed at the LHC with early data 
is discussed in Sec.~\ref{sec:hfcinttbar}. Conclusions are summarized in Sec.~\ref{sec:conclusion}.

\section{Top quark decays: Standard Model and beyond}
%\subsection{Experimental signature of top quarks}
\label{subsec:topexperimentalsignature}

Due to their large mass ($m_t > m_W+m_b$), top quarks decay promptly without hadronizing, $t\rightarrow Wb$.
Other decays, such as $t\rightarrow Ws$ and $t\rightarrow Wd$ are suppressed in the context of the SM, approximately
by a factor of $10^{-3}\div 10^{-4}$.
In the prompt decay scenario, the polarization of the top quark is 
transferred to the decay products and three final polarization states can be expected.
%The branching fractions for each polarization state are usually represented by $F_L$, $F_R$ and $F_0$ 
%corresponding to left/right transverse polarizations or longitudinal polarization of the $W$ boson correspondingly.
%There is no preferential direction for the emission of the $W$ boson and the top quark decay is isotropic (like an S-wave).
%The polarization acquired by the $W$ boson is therefore determined by the ratio of the $b$ and $W$ masses with respect to the Top mass.
As $m_b\ll m_t$, it is possible to assume in first approximation 
that the bottom quark behaves almost as a massless particle and that it can mostly acquire left-hand polarization, 
and the polarization of the top quark is transferred to the $W$ according to the ratio $m_W/m_t$.
In this approximation, right-polarized $W$ states are highly suppressed and the transversal polarization states are dominant. % ($\approx 70\%$). 

The $W$ boson decay channels determine the experimental signature of a $t\bar{t}$ event:
``fully hadronic'', ``lepton+jets'', or ``dilepton'' events accompanied by the presence of two $b$ jets are therefore 
expected. In hadron collisions,
$t\bar{t}$ pairs are also accompanied  by an underlying event which is the result of the proton remnants
and by extra {\it initial} and {\it final} state radiation (ISR/FSR) jets.

%Table~\ref{tab:ttbarchannels} summarizes the different final states expected for a $t\bar{t}$ pair.
%\begin{table}[htp]
%\centering
%\caption{Decay modes for a $t\bar{t}$ pair in the different channels and corresponding branching fractions in the context of the SM.
%$l=e,\mu$ represents the light charged leptons and $\tau$ represents the hadronic decaying $\tau$ leptons. 
%The last column shows the ``effective'' branching fractions when the $W$ and $\tau$ experimental values from~\cite{Amsler:2008zzb} are used 
%and the cross contributions from $\tau$ leptonic decays are included in channels with $e$ or $\mu$ in the final state.}
%\label{tab:ttbarchannels}
%\begin{tabular}{lccc} \hline
%Channel & Decay mode & \multicolumn{2}{c}{BR} \\\hline\hline
%~~Full hadronic~~ & $(q\bar{q}' b)~(q'\bar{q}'b)$~~ & 36/81 & ~~0.454 $\pm$ 0.004~~ \\\hline
%~~Lepton + Jets~~ & $(q\bar{q}' b)~(l\nu_l b)$~~ & 24/81 & ~~0.344 $\pm$ 0.003~~ \\\hline
%~~$\tau$ + Jets~~ & $(q\bar{q}' b)~(\tau\nu_\tau b)$~~ & 12/81 & ~~0.096 $\pm$ 0.002~~ \\\hline
%~~Dilepton~~ & $(l\nu_l b)~(l\nu_l b)$~~ & 4/81 & ~~0.065 $\pm$ 0.001~~ \\\hline
%\multirow{2}{*}{~~$\tau$-Dilepton~~} & $(\tau\nu_\tau b)~(l\nu_l b)$~~ & 4/81 & ~~0.036 $\pm$ 0.001~~ \\\cline{2-4}
%& $(\tau\nu_\tau b)~(\tau\nu_\tau b)$~~ & 1/81 & ~~0.0050 $\pm$ 0.0002~~ \\\hline
%\end{tabular}
%\end{table}

Among all, the dilepton channel has a particularly clean final state with
%as not many processes yield 
two oppositely charged prompt and isolated leptons with large $p_{T}$, at least two jets from $b$-decays and large missing transverse energy from the neutrinos.
The branching fraction of the $t\bar{t}$ dilepton channel is $B[t\bar{t}\rightarrow (l\nu_l b)(l\nu_l b)]=$~0.065$\pm$0.001 ($l=e,\mu$)~\cite{Amsler:2008zzb}.
SM processes that can mimic the dilepton channel involve mainly the production of $Z$ and $W$ bosons,
either as single particles or in pairs (i.e. di-boson production). 
For instance the same-flavor dilepton channels (i.e. $ee$ and $\mu\mu$) can be mimicked by
Drell-Yan processes, mainly from $Z$ boson production, 
that can be distinguished from $t\bar{t}$ events by excluding dileptons with a mass close to the $Z$ boson.
Single top production, in particular the $tW$ channel can also mimic the dilepton channel final state, but with a much lower cross section.
QCD events, although expected to be the dominant production at a hadron collider, are not expected to produce isolated, high-$p_T$ leptons.
Therefore the possible contamination from this processes is expected to be small, and mostly due to instrumental and misreconstruction effects.
In the dilepton channel the presence of at least two neutrinos accounts for an experimental transverse energy imbalance in the detector
which can possibly discriminate further $t\bar{t}$ events against background.

%\subsection{Beyond the SM signatures mimicking $t\bar{t}$}
%\label{subsec:bsmsignatures}

Processes not contemplated in the SM may also contaminate the $t\bar{t}$ sample. 
Searching for deviations of the basic properties of $t\bar{t}$ events, as they are predicted by the SM, 
is therefore an important topic of investigation at the LHC.
Measuring the production cross section, the mass, width and branching fractions of the top quark, 
the polarization of the final state particles and
the jet content of the events are some examples of these studies.
In this manuscript an emphasis is put on the search for deviations in the jet flavor content of the data sample.

Some examples which may alter the heavy flavor content of the $t\bar{t}$ sample can be found in the literature:
\begin{itemize}
\item associated Higgs boson production, decaying to a $b\bar{b}$ pair (i.e. $t\bar{t}H\rightarrow t\bar{t}b\bar b$).
Although the cross section is expected to be small, it may be enhanced
by non-SM processes such as 
$G'\rightarrow t'\bar{t}\rightarrow t\bar{t}H$
where $G'$ is a heavy color octet boson and $t'$ is a vector-like fermion~\cite{Dobrescu:2009vz}.
At the time this manuscript is written a lower mass limit of 311~GeV/$c^2$ has been set for $t'$ by the CDF 
experiment with 2.8~$fb^{-1}$ of data~\cite{Cox:2009mx};
\item if a $4^{th}$ generation of quarks exists, then decay cascades 
containing top quarks could be produced. These events are expected to have
either extra jets or additional leptons from the decays of multiple $W$ bosons.
Results from the CDF experiment set a lower limit for $m_{b'}>$325~GeV/c$^2$ with 2.7~fb$^{-1}$ of data~\cite{CDF:2009bp};
\item in the context of a supersymmetric theory, the mixing of chiral and scalar states 
could be privileged in the top quark sector due to its large mass.
If $\tilde{\chi}^0$ is assumed as the lightest supersymmetric particle
and the stop mass is $(m_{\tilde{\chi}^{+}_1}+m_b) < m_{\tilde{t}}\lesssim m_t$,
then $\tilde{g}\tilde{g}\rightarrow t\tilde{t}q\tilde{q}$ (or $\tilde{t}\tilde{t}$) is allowed
with the stop decaying further in $\tilde{t}\rightarrow t\tilde{\chi}^0$
is expected to have the typical signature of a $t\bar{t}$ pair plus
additional $\cancel{E}_T$ due to undetected neutral particles, and possibly larger lepton/jet multiplicities.
Limits on stop mass were obtained by the CDF experiment with 2.7~fb$^{-1}$ of data
for different values of $B(\tilde{\chi}^{+}_{1}\rightarrow \tilde{\chi}^0_1 l\nu_l)$ and 
$\chi^0_1$ masses~\cite{CDF:2009st};
\item the production of a charged Higgs from a top quark decay can also be expected in the
context of supersymmetry if $m_t>m_{H^{+}}+m_b$.
This scenario would be reflected by an increase in the production of the $\tau$ dilepton channel
with respect to the other channels from $H^{+}\rightarrow \tau\nu_\tau$ decays.
However, in the case of leptophobic $H^{+}$ an increase of charm and strange particles 
from the production of $c\bar{s}$ pairs could also be expected.
With 1.0~fb$^{-1}$ the D0 collaboration performs a model-independent
measurement which excludes $B(t\rightarrow H^+b)>$~0.12-0.26 at 95\% C.L.
depending on $M_{H^+}$\cite{:2009zh}.
\end{itemize}

Using a different approach from the various phenomenological scenarios just briefly reviewed, 
a generic model-independent search aimed at the study of
the heavy flavor content of $t\bar{t}$ events is proposed in the following~\cite{Psilva:2009th}.
If measurements in the data are compatible with the SM expectations,
$V_{tb}$ can be determined under the particular assumption on the structure of the CKM matrix.

\section{Probing the heavy flavor in $t\bar{t}$ events}
\label{sec:hfcinttbar}

Decay products of top quark events are expected to contain $b$-jets,
resulting either from the top quarks or other heavy particle decays as reviewed earlier.
Experimentally, the flavor of a jet can be identified by taking advantage of the properties of the hadrons containing $b$ quarks.

$B$ hadrons are long-lived with proper-lifetimes typically of the order of $c\tau_{B}\approx$~450~$\mu m$. %~\cite{Amsler:2008zzb}.
After the decay of long-lived hadrons into charged particles, %e.g. BR$(B^0\rightarrow K^\pm +X)~\sim$~78\%,
a secondary vertex of tracks associated to a jet might be found.
In $\approx$20\% of the cases these jets can also contain an electron or a muon due to the semi-leptonic branching ratio of the $B$ mesons.
Therefore, jets produced by heavy flavor quarks can be identified by displaced tracks or secondary charged leptons.
The ``$b$-tagging'' algorithms exploit these simple properties to build discriminator variables which can be used to identify the flavor of a jet. 
Each algorithm is usually characterized by an efficiency of identifying $b$ jets ($\varepsilon_b$) 
and a ``mis-tag'' rate due to mis-identification of light quark and gluon initiated jets ($\varepsilon_q$)~\cite{CMSPas:BTV0901}.
Experimentally, $\varepsilon_b$ can be measured in QCD events with reconstructed jets containing muons. 
These events are most likely $b$-jet enriched events (i.e. from $b\bar{b}$) 
and allow one to evaluate the $b$-tagging efficiency using a ``tag and probe'' method.
The mistag rate can also be estimated from data using 
the fact that for light quark and gluon-initiated jets the tracks are distributed
randomly in the vicinity of the primary vertex of the hard interaction.
Therefore, the tracks produced upstream with respect to the primary vertex of the hard interaction
can be used to extrapolate the contamination in the $b$-tagging region. 
Both measurements are sensitive to the alignment of the tracker detector,
to the track reconstruction efficiency and resolution of the track impact parameter
and to the resolution of the vertices.
%and to the resolution in the reconstruction of the lepton impact parameter (if electrons or muons are used as ``$b$ taggers'').

By using the $b$ tagging algorithms just described, a method can be developed to probe flavor deviations
in the decays of $t\bar{t}$ pairs.
Generator level simulations are used in the following to illustrate how to apply the method to the data.

\subsection{Description of the method}
\label{subsec:method}

\begin{figure}[htp]
\centering
\includegraphics[width=0.49\textwidth]{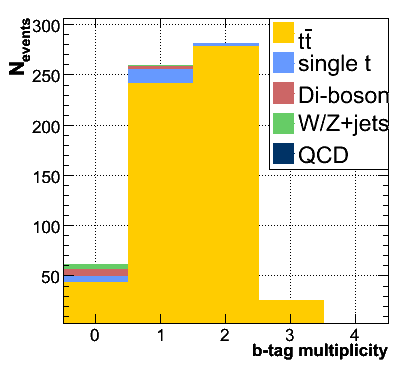} \hfill 
\includegraphics[width=0.49\textwidth]{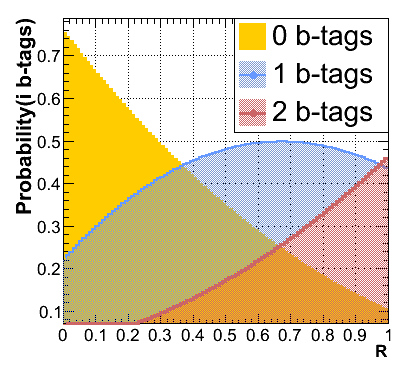} \hfill 
\caption{({\em Left}) $b$-tagging multiplicity obtained assuming a jet is $b$-tagged assuming $\varepsilon_b$=80\% and $\varepsilon_q$=15\%.
({\em Right}) Probability model for the observation of $k$ $b$-tags in events with two jets, as function of $R$.}
\label{fig:btagmultiplicity}
\end{figure}

Evidence for the heavy flavor content of the $t\bar{t}$ final states can be obtained by counting the $b$-tag multiplicity found in the selected events.
The number of $b$-tags counted in the events is proportional to the fraction of $b$~jets present in the sample, and is related to 
$\varepsilon_b$ and $\varepsilon_q$.
Ultimately, the number of $b$~jets produced depends on the ratio of branching fractions of the top quark to bottom quarks
(Eq.~\ref{eq:r}).
In particular, a simple selection of $t\bar{t}$ dilepton events can be devised
where the final sample is dominated by top quark events over the background ($S/B\approx 16$).
An example of the $b$-tag multiplicity distribution of $t\bar{t}$ dilepton events is shown in Fig.~\ref{fig:btagmultiplicity} ({\em left}).
The distribution is obtained from Monte Carlo (MC) using typical values for $\varepsilon_b$ and $\varepsilon_q$, 
randomly thrown to assign each jet a $b$-tag. 
The two $b$-tag multiplicity bin clearly dominates, as expected in a $t\bar{t}$-dominated sample where two jets are from heavy flavor jets.

Although in the dilepton channel small contributions are expected from background processes,
there is a non-negligible probability that at least one jet from a $t\bar{t}$ decay is either missed because it was not reconstructed 
or because it did not pass the jet selection criteria. 
Another jet that may be selected instead could, for example, be a jet initiated from ISR or FSR.
This will be referred to as ``jet misassignment'' and its determination from data is discussed in detail in Sec.~\ref{subsec:jetmisassignment}. 
The b-tag multiplicity distribution reflects also the fraction of jet misassignment in the sample,
namely the fractions of events with:

\begin{itemize}
\item no jet from the top decays selected (background-dominated);
\item only one jet correctly assigned to a top decay (combination of signal and background);
\item two jets correctly assigned to the top decays (signal-dominated).
\end{itemize}

The relative contributions of these three classes of events can be expressed by the
weights $\alpha_{i}$, where $\sum_{i}\alpha_{i}=1$ and $i=0,1,2$ is the number of jets from top decays correctly reconstructed and selected. 
For example, $\alpha_2$ is the probability that both jets are correctly assigned to $b$ jets.

Using $R$, $\varepsilon_b$, $\varepsilon_q$ and the $\alpha_i$ (as previously defined),
it is possible to model probabilistically the average $b$-tagging multiplicity.
The number of expected events with $k$~tagged jets can be generically written as follows:

\begin{equation}
\hat{N}_{ev}(k~b-tags) =\sum_{\text{n~jets}=2}^{\text{all jets}} N_{ev}(n~jets) \cdot P^{n~jets}_{k~b-tags}
\label{eq:btagmultmodel}
\end{equation}

where $N_{ev}(n~jets)$ is the number of observed events with $n$-jets, and $P^{n~jets}_{k~b-tags}$ 
is the probability to count $k$ b-tags in a $n$-jet event.
The probability function encloses the available knowledge on $R$, $\varepsilon_b$, $\varepsilon_q$ and $\alpha_i$.
In order to illustrate explicitly the construction of such probability functions, the exclusive two-jet multiplicity bin 
is used and the following expression is obtained:

\begin{equation}
P^{2~jets}_{k~b-tags} = \mathop{\sum_{\text{i jets}=0}^{2}}_{\text{from~top~decay}} \alpha_{i}\cdot P^{~i~|~2~jets}_{k~b-tags}
\label{eq:twojetsprobmodel}
\end{equation}

where the $\alpha_i$ parameters describe the composition of the sample in terms of events with $i$-jets from top quark decays
correctly reconstructed and selected, and $P^{~i~|~2~jets}_{k~b-tags}$ is the probability that $k$~$b$-tags are observed in an event 
with two jets of which $i$~jets come from $t\bar{t}$ decays.
Extra jets can be reconstructed and selected in the selected events mainly coming from Initial/Final state radiation
produced along with the $t\bar{t}$ pair. This extra jets can mimic $t\bar{t}$ events even if not coming directly from top decays.
The previous expression can therefore be further developed and, 
in particular for the case where two tagged jets are found in a 2-jet event, 
the probability can be explicitly written as: 

\begin{equation}
\begin{array}{lll}
P^{~0~|~2~jets}_{2~b-tags} = \varepsilon_q^2  & 
 & \text{\small if no jets are from $t\bar{t}$ decays}\\
& & \\
P^{~1~|~2~jets}_{2~b-tags} = 2R^2\varepsilon_b\varepsilon_q+2R(1-R)(\varepsilon_b+\varepsilon_q)\varepsilon_q+2(1-R)^2\varepsilon_q^2  & 
 & \text{\small if 1 jet is from $t\bar{t}$ decays}\\
& & \\
P^{~2~|~2~jets}_{2~b-tags} = R^2\varepsilon_b^2+2R(1-R)\varepsilon_b\varepsilon_q+(1-R)^2\varepsilon_q^2 & 
 & \text{\small if 2 jets are from $t\bar{t}$ decays}
\end{array}
\label{eq:twotagstwojetbinprob}
\end{equation}

So far, the probability model has been derived for the most relevant case which is the observation
of two $b$-tags in the 2-jet multiplicity bin. The model can be expanded to a more general situation 
but it is omitted from this manuscript for the sake of simplicity.
%Expanding the model, for other cases is done similarly to the example just described and it is omitted for simplicity.
In this probability model it is assumed that:

\begin{enumerate}
\item the correlation between the probability of observing single and double tags is negligible;
\item for higher jet multiplicity bins the probability can be computed from the product of $P^{2~jets}_{k~b-tags}$ 
  with the probability of observing the extra tags from the extra radiation jets;
\item the mistag rate $\varepsilon_q$ is an effective measurement of the probability of ``tagging'' light quark, gluon and charmed jets
  in the dilepton sample. In first approach, it is taken as the mistag rate for light quark and gluon jets but can also be derived 
  directly from data as it will be shown in the next Section.
\end{enumerate}

By summing up the contributions from events with 2, 1 or no correctly assigned jets (weighted by $\alpha_{i}$), 
as expressed in Eq.~\ref{eq:twojetsprobmodel},
one builds the probability model shown in Fig.~\ref{fig:btagmultiplicity} ({\em right}).

Using the generic Eq.~\ref{eq:btagmultmodel}, $R$ can be fit from the $b$-tag multiplicity distributions
by maximizing the likelihood function:

\begin{equation}
{\mathcal L}(R,\varepsilon_{b},\varepsilon_{q},\alpha_{i}) = \prod_{k=0}^{all~jets}
{\mathcal Poisson} [N_{ev}(k~b-tags), \hat{N}_{ev}(k~b-tags)]
\label{eq:hfcfitlikelihood}
\end{equation}

where $N_{ev}(k~b-tags)$ is the number of observed events with $k~b-$tags.
Equation~\ref{eq:hfcfitlikelihood} can be used to perform the fit using different choices and sub-samples of the selected events.
The results obtained will be discussed in more detail by applying the model to a MC generated sample.

\subsection{Dilepton sample selection}
\label{sec:sampleselection}

In order to illustrate the potential of the method described, a dilepton sample was generated
using MadGraph~\cite{Maltoni:2002qb} and Pythia~\cite{Sjostrand:2006za} generators with $R=1$.
Inclusive $t\bar{t}$ events and the most relevant background processes: single top, W/Z+jets, WW, WZ, ZZ and QCD were generated.
The events are selected mimicking a realistic dilepton event selection from reconstructed events
as the ones proposed by the ATLAS and CMS collaborations~\cite{Aad:2009wy,CMSPas2009:top01,CMSPas2009:top02}.

%Events are ``triggered'' by at least a single electron or muon.
% with transverse momentum $p_T>$15~GeV/c and pseudo-rapidity $|\eta|<$3.0.
A simple event selection requires the presence of two high-$p_T$ isolated leptons, either electron or muon, at least two jets, and large missing transverse energy.
This simple selection of the $t\bar t$ dilepton channel yields a clean sample with $S/B\approx 16$
\footnote{For proton-proton collisions at a center of mass energy of 7 TeV 
it can be inferred from the re-scaling of signal and background cross sections
obtained with the Pythia generator that the S/B figure of merit should decrease by approximately 30\%.
It must be noted that cuts have been gauged for 10 TeV collisions and optimization 
of the S/B can be performed for 7 TeV collision.}. 
For the study presented here it is expected to collect 650-700 events, whereas only 6\% 
of the events are from background sources 
when an integrated luminosity of 100~pb$^{-1}$ at $\sqrt{s}$=10~TeV is considered.
The exact number of the events selected is not relevant to the results of the method presented here, as it will become clear.

In the selected samples (signal and background), the number of events with 2, 1 or 0 jets 
from top quark decays correctly reconstructed and selected can be counted using MC truth.
Table~\ref{tab:alphaimctruth} shows the fraction of events with 2, 1 or 0 jets
(i.e. the values for the $\alpha_i$ introduced in Sec.~\ref{subsec:method}).
In first approximation, the $\alpha_i$ can be parameterized as a binomial combination of $\alpha$, 
the probability that a jet from a top quark decay is reconstructed and selected. 
The probability to have two jets from $t\bar{t}$ decays reconstructed and correctly assigned (independently of its flavor) is therefore given by $\alpha_2=\alpha^2$. 
Conversely, the probability that both jets from the top decay are missed in the selected event is given by $\alpha_0=(1-\alpha)^2$. 
One jet from top decay is expected to be selected with a probability given by $\alpha_1=2(1-\alpha)\alpha$.
The values for $\alpha$ are also shown in Tab.~\ref{tab:alphaimctruth} and were obtained by fitting a binomial to the $\alpha_i$.

\begin{table}[htp]
\caption{Values of $\alpha_i$ for the selected events computed using MC truth. Different dilepton channels are shown
The last column shows the value of $\alpha$ that fits best the $\alpha_i$ with a binomial function (see text).
Only statistical uncertainties are shown.}
\label{tab:alphaimctruth}
\begin{center}
\hspace*{-0.5cm}
\begin{tabular}{c|ccc|c}\hline
Dilepton channel & $\alpha_2$ & $\alpha_1$ & $\alpha_0$ & $\alpha$ \\ \hline \hline
$e\mu$      & 0.675 $\pm$ 0.042 & 0.295 $\pm$ 0.024 & 0.030 $\pm$ 0.005 & 0.824 $\pm$ 0.009 \\ \hline
$ee/\mu\mu$ & 0.635 $\pm$ 0.041 & 0.288 $\pm$ 0.023 & 0.077 $\pm$ 0.009 & 0.78 $\pm$ 0.01 \\ \hline
\end{tabular}
\end{center}
\end{table}

%In the next Section the determination of $\alpha_i$ from data is discussed.
%Determining these values directly from data is essential to reduce the uncertainties due to MC simulation.
%when interpreting real data for the purposes of measuring the top quark decay in $t\bar{t}$ events.

How can the correct values of $\alpha_i$ be determined? It will be explained in the following.

\subsection{Jet misassignment estimate from data}
\label{subsec:jetmisassignment}

The $\alpha_i$ values can be estimated directly from data using the kinematic properties of the events.
The method presented suggests to minimally rely on the MC simulation in order to minimize the uncertainty due to the MC model.

%The probability data model used to probe the heavy flavor content of the $t\bar{t}$ sample
%relies on the knowledge of the jet assignment probabilities, i.e. $\alpha_i$, 
%the probability that $i$-jets from top decays have been reconstructed and selected.
%The $\alpha_i$ can be estimated directly from data using the kinematic properties of the events.

The lepton-jet pairs originating from the same top quark decay are kinematically correlated.
The distribution of the angle between the lepton and jet directions of motion in the $W$-boson rest frame,
and the lepton-jet invariant mass are often pointed out as possible variables which depict experimentally 
the existing correlation between particles produced from the same top decay cascade~\cite{Kane:1991bg}.
The lepton-jet invariant mass shape is Jacobian with a kinematical end-point 
at $M_{l,b}^{max}\equiv\sqrt{m_{t}^{2}-m_{W}^{2}}\approx$156~GeV/c$^{2}$ 
after which no lepton-jet pair from $t\bar{t}$ decays is observed.
On the other hand, the invariant mass of mis-assigned lepton-jet pairs exhibits 
a longer tail towards high mass values. The shape of these distributions depend on the $t\bar{t}$ pair kinematics
and on the contamination from ISR/FSR jets in the sample.
Two methods are proposed to model the invariant mass distribution of the mis-assigned lepton-jet pairs (i.e. background model):

\begin{itemize}
\item{\bf ``random rotation''} of the momentum of the selected leptons, $\vec{p}\rightarrow p\vec{e}(\theta,\phi)$;
\item{\bf ``swap''} the jet in the assigned lepton-jet pair, with a jet from a different event.
\end{itemize}

\begin{figure}[htp]
\centering
\includegraphics[width=0.49\textwidth]{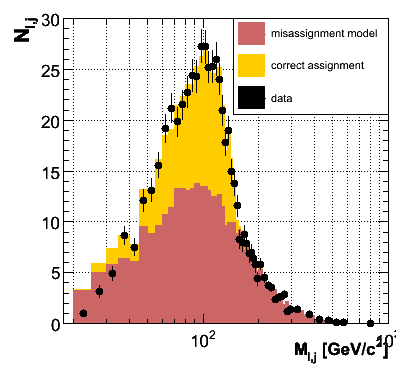}\hfill
\includegraphics[width=0.49\textwidth]{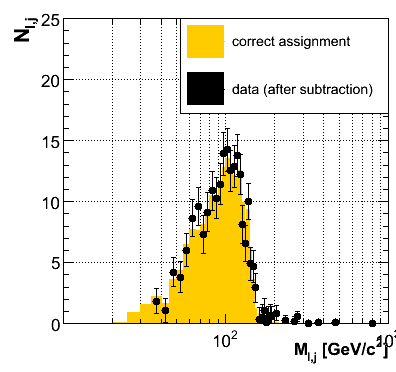}
\caption{ 
({\em Left}) Inclusive lepton-jet invariant mass distribution. The spectrum obtained from the mis-assignment model
described in the text is shown together with the distribution of the correctly (using MC truth) assigned lepton-jet pairs.
({\em Right}) Invariant mass distribution of the lepton-jet pairs after the background is subtracted.}
\label{fig:mlj}
\end{figure}

In practice these two methods yield similar invariant mass spectra and their average is used,
which will be referred to as the combinatorial ``mis-assignment" model.
Figure~\ref{fig:mlj} ({\em left}) shows the invariant mass distribution of the mis-assignment model
after it has been rescaled in order to reproduce the tail of the inclusive lepton-jet spectrum for masses $M_{l,j}>$190~GeV/c$^{2}$.
In the ``data'' sample, each lepton (electron or muon) is paired with all selected jets and the invariant mass is computed for all pairs.
The distribution of the correctly assigned lepton-jet pairs, derived from MC truth information, is also shown.
%stacked on top of the misassignment model.
The correct assignment distribution together with the mis-assignment model sums up correctly (within uncertainties)
reproducing the inclusive lepton-jet invariant mass spectrum (``data'' points).
The mis-assignment model describes well the expected ({\it true}) lepton-jet mis-assignment
distribution in most of the mass region of the spectrum ($M_{l,j}>40$~GeV/c$^{2}$).
The low mass region of the spectrum ($M_{l,j}<40$~GeV/c$^{2}$),
which is due to almost collinear lepton-jet pairs, is expected to be suppressed in real 
data because the leptons are required to be isolated and the jets to be separated from the selected leptons~\cite{Aad:2009wy,CMSPas2009:top02}.

Figure~\ref{fig:mlj} ({\em right}) shows the distributions of the lepton-jet invariant mass pairs after 
the subtraction of the mis-assignment model.
Good agreement with the correct assignment distribution (obtained using MC truth) is found in the high mass region ($M_{lj}>190$~GeV/c$^{2}$)
where the mis-assignments dominate and no signal lepton-jet pairs are expected.

The contribution of jet mis-assignment (i.e. background) can be fit directly from the mis-assignment model
by computing the ratio of areas below the $M_{l,j}$ distributions, in ``data'' and in the mis-assignment model'' (Fig.~\ref{fig:mlj}, left).
The ratio is expected to flatten out in the region dominated by the background 
(typically $M_{l,j}>$190~GeV/c$^2$) and it provides the normalization factor of the background.
If:

\begin{itemize}
\item { $n(M_{l,j})$ is the distribution of all lepton-jet pairs as a function of the pair mass (total number of pairs is $N$);}
\item { $n_{mis}(M_{l,j})$ is the distribution of only mis-assigned pairs (total number of mis-assignments is $N_{mis}$)},
\end{itemize}

then, for $M_{l,j}>z$, it is assumed that all the pairs are mis-assigned.

\begin{equation}
N_{mis}^{M_{l,j}>z} = N^{M_{l,j}>z}
\label{eq:misnormeq}
\end{equation}

The fraction-$f$ of background pairs with $M_{l,j}>z$ can be used to estimate the total number of mis-assignments, $N_{mis}$.
If $N_{model}^{M_{l,j}>z}$ is the number of background pairs predicted by the background model with  $M_{l,j}>z$
and $N_{model}$ is normalized to the total number of pairs, then it is assumed that:

\begin{equation}
f=\frac{N_{mis}^{M_{l,j}>z}}{N_{mis}} = \frac{N_{model}^{M_{l,j}>z}}{N_{model}}
\label{eq:mismodelcondeq}
\end{equation}

Using the previous result, one can normalize the background model to the signal depleted range (i.e.$M_{l,j}>z$)
Then, the fraction of correctly assigned lepton-jet pairs is given by:

\begin{equation}
\alpha = \frac{N-N_{mis}}{2~N_{evts}} = \frac{N-\frac{N_{mis}^{M_{l,j}>z}}{f}}{2~N_{evts}} = 
\frac{N}{2~N_{evts}}\left(1-\frac{N_{mis}^{M_{l,j}>z}}{N_{model}^{M_{l,j}>z}} \right)
\label{eq:misassignmentfrac}
\end{equation}

The value of $\alpha$, estimated by applying Eq.~\ref{eq:misassignmentfrac} to the lepton-jet invariant mass spectrum,
yields $\alpha=$ 0.82 $\pm$ 0.03 for the ``$e\mu$'' sample,
and $\alpha=$ 0.80 $\pm$ 0.03 for the ``$ee$ and $\mu\mu$'' sample,
in agreement with the MC-expected values of Tab.~\ref{tab:alphaimctruth} (right column), within uncertainties.
%The fraction of jet mis-assignments extracted from ``data'' is in good agreement with MC truth within uncertainties.

%\subsubsection{Systematic uncertainties}
%\label{subsubsec:alphasysts}

\begin{figure}[htp]
\begin{center}
\includegraphics[width=0.49\textwidth]{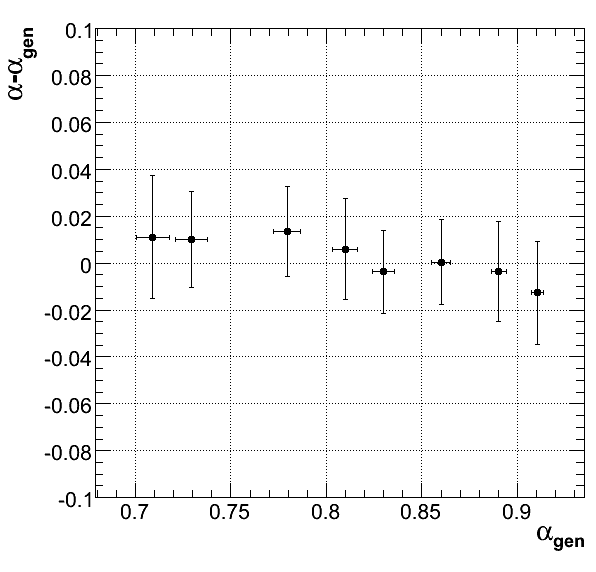}
\caption{Deviation of the value of $\alpha$ derived from ``data'' with respect to the MC value, $\alpha_{gen}$ (see text).}
\label{fig:alphapull}
\end{center}
\end{figure}

The robustness of the method described for the determination of $\alpha$ can be tested by artificially modifying some of the parameters.
The selected events are re-weighted depending on the number of jets from top decays, after reconstruction and selection.
The resulting distribution of the difference between the value obtained from ``data'' %with the method
and the true generated value, i.e. $(\alpha-\alpha_{MC})$, is shown as a function of $\alpha_{MC}$ in Fig.~\ref{fig:alphapull}.
The results are compatible with zero bias, meaning the method is able to reproduce the correct number of reconstructed jets from top quark decays.
Variation of other parameters which are affected by experimental uncertainties,
such as lepton and jet energy scales, or which are intrinsic to the method itself,
such as the mass cut used to determine the normalization of the misassignment model,
are are listed in Tab.~\ref{tab:alphasystematics}.
Overall, these variations account for $<$0.02\% in the determination of $\alpha$.
One can therefore conclude that the determination of $\alpha$ 
is expected to robust and to be dominated by the statistical uncertainty.

\begin{table}[htp]
\caption{Deviations in $\alpha$ from the variation (shift) of different parameters.
The determination of $\alpha$ was repeated after varying independently each parameter presented in the left column of the table (see text).
Variations are presented separately for each channel.}
\label{tab:alphasystematics}
\begin{center}
\begin{tabular}{lcc}\hline
\multirow{2}{*}{Parameter} & \multicolumn{2}{c}{Channel} \\ 
 & $e\mu$ & $ee/\mu\mu$ \\ \hline \hline
{\qquad \small mass cut (5\%)}                & $^{+0.005}_{-0.01}$         & $^{+0.006}_{-0.01}$ \\ \hline
{\qquad \small JES (10\%)}                    & $^{+0.01}_{-0.01}$         & $^{+0.01}_{-0.01}$ \\ \hline
{\qquad \small LES (1\%)}                     & $^{+0.006}_{-0.006}$        & $^{+0.009}_{-0.009}$ \\ \hline
Total systematic (quad sum)                   & {\large $^{+0.01}_{-0.02}$} & {\large $^{+0.01}_{-0.02}$} \\ \hline
\end{tabular}
\end{center}
\end{table}

\subsection{Measurement of R}
\label{subsec:rmeasurement}

\begin{figure}[htp]
\begin{center}
\includegraphics[width=0.49\textwidth]{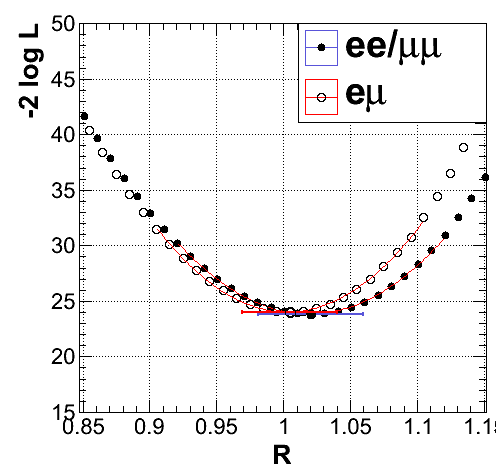}\hfill
\includegraphics[width=0.49\textwidth]{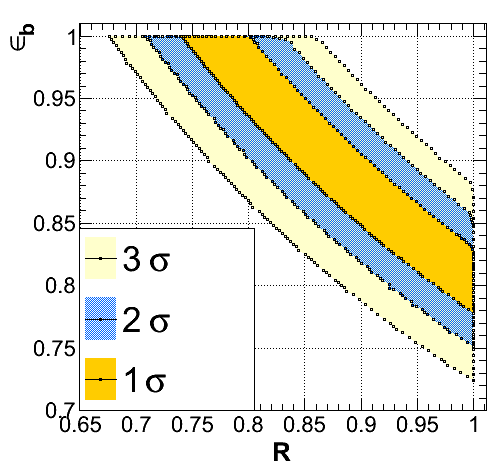}
\caption{({\em Left}) Fit results for $R$.
({\em Right}) Contour plots (1$\sigma$, 2$\sigma$ and 3$\sigma$)  for the likelihood obtained by floating R and $\varepsilon_b$.
In the MC, $R=$1, $\varepsilon_b=$0.80, $\varepsilon_q=$0.15 for both plots.}
\label{fig:rfit}
\end{center}
\end{figure}

The measurement of $R$ can be performed by fitting the $b$-tagging multiplicity distribution (Fig.~\ref{fig:btagmultiplicity}) 
using the probability model (Eq.~\ref{eq:hfcfitlikelihood}):
$R$ is let to float freely without any constraint, and $\varepsilon_b$ and $\varepsilon_q$ are taken as inputs.
The $\alpha_i$ are measured from the ``data'' as described in Sec.~\ref{subsec:jetmisassignment}.
Figure~\ref{fig:rfit} ({\em left}) shows the results obtained by fitting $R$ to the $b$-tagging multiplicity
distribution for $ee/e\mu$ and $e\mu$ events, separately.
The result is statistically compatible with MC expectations: $R=$ 1.00 $\pm$ 0.03 (1.02 $\pm$ 0.04) for the $e\mu$ ($ee/\mu\mu$) sample.
By performing several pseudo-experiments and repeating the fit to the $b$-tagging multiplicity distributions
one obtains a good agreement with the MC truth, i.e.
$\delta R=<R>-R^{MC}=$~ -0.006 $\pm$ 0.04 ( -0.008$\pm$ 0.04 ) for the $e\mu$ ($ee/\mu\mu$) sample.
The procedure has been also tested on pseudo-experiments where $R^{MC}<1$. 
The fit converges correctly to the generated value with similar uncertainties as the ones quoted above.

The method can be easily extended to the simultaneous measurement of $R$ and other quantities, such as
$\alpha_0$ (i.e. the contribution from events where no jet from top quark decays has been reconstructed and selected),
$\varepsilon_q$ (i.e. the mistagging efficiency), or the $t\bar{t}$ cross section
which is discussed in more detail in Sec.~\ref{subsec:crosssection}.
Such fits can be used as a cross check for the value of $\alpha_0$ or $\varepsilon_q$ used as inputs for the measurement of $R$,
or in the latter case to perform a simultaneous measurement of two variables.
It is important to notice that the convergence of the fit is mostly driven by the experimental knowledge of $\varepsilon_b$.
As $R$ and $\varepsilon_b$ are correlated it is not possible to measure simultaneously these quantities from the same sample. 
Figure~\ref{fig:rfit} ({\em right}) shows the contour plots of $R$ and $\varepsilon_b$.
This contour plot can also be used to cross check the $b$-tagging efficiency used as input to the fit.

The robustness of the method used to determine $R$ is dominated by the uncertainty on the knowledge of the $b$-tagging efficiency.
With the first data acquired and the events selected for this measurement, only an average value for $\varepsilon_b$ can be used due to the limited number of events.
When more data will be available, the jet $p_T$ and $\eta$ dependence of $\varepsilon_b$ can be used.
The measurement of $R$ is also affected by the variation in $\varepsilon_q$ and $\alpha$.
The uncertainty in $\alpha$ is affected by the factors discussed in Sec.~\ref{subsec:jetmisassignment}.
These uncertainties are small when compared to the uncertainty from the $b$-tagging efficiency.
Variation of ISR/FSR contamination of the sample does not affect much the final result of the fit
as these variations are correctly described when $\alpha$ is estimated from data. 
Table~\ref{tab:rfitsysts} summarizes how $R$ is expected to be affected from the variation of the different fit parameters.

\begin{table}[htp]
\caption{Variation of $R$ due to the variation of the different fit parameters.
$\varepsilon_b$ is varied by $10\%$ and $\varepsilon_q$ is varied by $5\%$.
All other parameters are varied as discussed previously in Sec.~\ref{subsec:jetmisassignment}.}
\label{tab:rfitsysts}
\begin{center}
\begin{tabular}{lcc}\hline
Parameter                                  & $e\mu$               & $ee/\mu\mu$          \\ \hline\hline
{\qquad\small $\varepsilon_b$ (10\%)}      & $^{+0.10}_{-0.10}$   & $^{+0.10}_{-0.10}$   \\ \hline
{\qquad\small $\varepsilon_q$ (3\%)}       & $^{+0.02}_{-0.01}$   & $^{+0.02}_{-0.01}$   \\ \hline
{\qquad\small mass cut for $\alpha$ (5\%)} & $^{+0.03}_{-0.02}$   & $^{+0.03}_{-0.02}$   \\ \hline
{\qquad\small JES (10\%)}                  & $^{+0.02}_{-0.02}$   & $^{+0.02}_{-0.02}$   \\ \hline 
{\qquad\small LES (1\%)}                   & $^{+0.005}_{-0.005}$ & $^{+0.005}_{-0.005}$ \\ \hline
Total systematic                           & $\pm 0.1$            & $\pm 0.1$             \\ \hline
\end{tabular}
\end{center}
\end{table}

The dominant uncertainty is due to $\varepsilon_b$ (10\%) which is taken in the early
data-taking phase as an external input, and it is estimated from other physics processes.
The uncertainty in the measurement of $R$ is comparable to the reach of the Tevatron experiments at the time this manuscript was written.
With more data it is reasonable to expect this uncertainty to decrease significantly~\cite{Aad:2009wy,CMSbtag:2007pas}.
Notice that the parameters affect the different dilepton channels in the same way.
This feature is particularly important because it allows one to cancel out some
systematic uncertainties by taking the ratio of the measurements obtained in the different channels.
Combined measurements of $R$ can therefore be performed with the purpose of reducing 
the systematic uncertainties.
A procedure for combined measurements is proposed in the next Section.

\subsection{Combination of results in different channels}
\label{subsec:combresults}

The experimental uncertainty in the $b$-tagging efficiency is the dominant factor affecting the measurement of $R$.
%the heavy flavor content of $t\bar{t}$ events.
However this uncertainty can be controlled by combining the measurements of $R$ obtained from different channels.
The procedure can be in fact generalized. % for any unknown parameter entering the fit.
For instance, the uncertainty in $\alpha$ can be controlled by combining the measurements of $R$ 
obtained from different jet multiplicity bins in the same $t\bar{t}$ channel.
With early data an iterative procedure can therefore be adopted
taking advantage of different $t\bar{t}$ channels or control regions of the sample
in order to control the systematic uncertainties affecting this measurement.
Generically one can:

\begin{itemize}
\item divide the sample in two regions: A and B - e.g. two different channels or two different jet multiplicity bins;
\item fit the $b$-tag multiplicity distributions:
\begin{itemize}
\item in sample A, assume $R=1$ and leave the parameter $x=\varepsilon_b,\alpha,...$ to float freely in the fit;
\item in sample B, assume an initial estimate for parameter $x=\varepsilon_b,\alpha,...$ and leave $R$ to float freely in the fit;
\end{itemize}
\item the fit yields a value of $x$ for sample A and a value of $R$ for sample B;
\item repeat the fit giving as input $R(B)$/$x(A)$ to sample A/B;
\item if the value $R(B)$/$x(A)$ maximizes the likelihood in sample A/B it is kept otherwise the previous value is maintained;
\item the fit is repeated until no variation is observed in the values of $R$ and $x$ obtained from each sample.
\end{itemize}

As an example, such procedure is applied to region A (``$ee/\mu\mu$'' sample) and region B (``$e\mu$'' sample).
The $b$-tagging efficiency is measured in region A, and $R$ is measured in region B.
As $\varepsilon_b$ and $R$ are fully correlated, a Gaussian term is added to the likelihood 
presented in \ref{eq:hfcfitlikelihood} as follows:

\begin{equation}
{\mathcal L}(R,\varepsilon_{b},\varepsilon_{q},\alpha_{i}) = \prod_{k=0}^{all~jets} 
{\mathcal Poisson} [N_{ev}(k~b-tags), \hat{N}_{ev}(k~b-tags)]
\times Gaussian(\varepsilon_b,\bar{\varepsilon_b},\sigma_{\varepsilon_b})
\label{eq:hfcfitlikelihoodgausconstrain}
\end{equation}

where $\bar{\varepsilon_b}$ is the experimentally measured value of $\varepsilon_b$ in a $b$ jet enriched sample (see Sec.~\ref{sec:hfcinttbar})
and $\sigma_{\varepsilon_b}$ is the corresponding uncertainty.
This Gaussian extra term makes the fit more robust, as it takes into account the proper uncertainty associated to $\varepsilon_b$.
The final result obtained is compatible with MC truth:
$\delta \varepsilon_b=<\varepsilon_b>-\varepsilon_b^{MC}=$0.005~$\pm$~0.01 and $\delta R=<R>-R^{MC}=$-0.004~$\pm$~0.02.
Table~\ref{tab:rfitsystswithfb} summarizes how the variation of the different parameters
affects the measurement of $\varepsilon_b$ and $R$ in the different channels.
The initial uncertainty on the knowledge of $\varepsilon_b$ is
reduced by performing the proposed method.
%which is robust for the measurement of $R$.
This result is particularly interesting given the fact that
throughout this manuscript we have considered a sample corresponding to 100~pb$^{-1}$. 
Increased statistics and the knowledge of the detectors, together with additional inputs from other $t\bar{t}$ channels
will certainly reduce the total uncertainty in the measurement of this quantity.

\begin{table}[htp]
\caption{Variations of $R$ and $\varepsilon_b$ in the $e\mu$ and $ee/\mu\mu$ channels
when an iterative procedure is used to measure this quantities simultaneously (see text).}
\label{tab:rfitsystswithfb}
\begin{center}
\begin{tabular}{lcc}\hline
Sample                                     & $e\mu$                & $ee/\mu\mu$\\ \hline\hline
Parameter                                  & $R$                   & $\varepsilon_b$\\ \hline
{\qquad\small $\varepsilon_b$ (10\%)}      & $^{+0.004}_{-0.003}$  & $^{+0.003}_{-0.003}$\\ \hline
{\qquad\small $\varepsilon_q$ (3\%)}       & $^{+0.003}_{-0.001}$  & $^{+0.005}_{-0.005}$ \\ \hline
{\qquad\small mass cut for $\alpha$ (5\%)} & $^{+0.004}_{-0.004}$  & $^{+0.001}_{-0.001}$ \\ \hline
{\qquad\small JES (10\%)}                  & $^{+0.003}_{-0.003}$  & $^{+0.005}_{-0.005}$ \\ \hline
{\qquad\small LES (1\%)}                   & $^{+0.0007}_{-0.001}$ & $^{+0.003}_{-0.004}$ \\ \hline 
Total systematic                           & {\large $^{+0.007}_{-0.006}$} & {\large$^{+0.008}_{-0.009}$} \\ \hline
\end{tabular}
\end{center}
\end{table}

\subsection{Cross section measurement}
\label{subsec:crosssection}

\begin{figure}[htp]
\begin{center}
\includegraphics[width=0.6\textwidth]{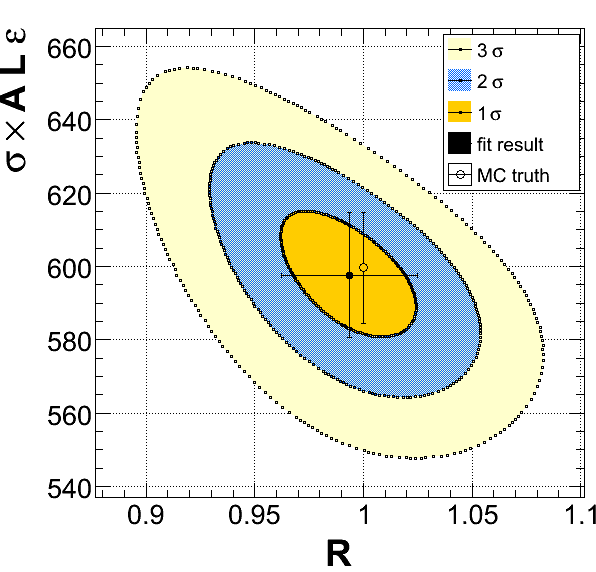}
\caption{Combined fit results for $R$ and $\sigma$ using the full dilepton sample.
Contour plots (1$\sigma$, 2$\sigma$ and 3$\sigma$)  for the likelihood obtained by floating $R$ and $\sigma$
are shown. In the MC, $R=$1, $\sigma=$414~pb.}
\label{fig:randcrosssectionfit}
\end{center}
\end{figure}

As stated in Sec.~\ref{subsec:rmeasurement} it is possible to extend the method of measuring $R$
to incorporate a simultaneous measurement of the cross section. 
The production cross section can be determined experimentally from the following expression:

\begin{equation}
\sigma = \frac{ N_{t\bar{t}~\text{dileptons}} }{A\cdot\varepsilon\cdot L}
\label{eq:crosssection}
\end{equation}

where  $N_{t\bar{t}~\text{dileptons}}$ 
is the total number of $t\bar{t}$ dilepton events above the background, 
$A$ is the acceptance of the $t\bar{t}$ dilepton channel,
$\varepsilon$ combines trigger, reconstruction and selection efficiencies,
and $L$ is the total integrated luminosity.
The determination of both $A$, $\varepsilon$ and $L$ is beyond the scope of this manuscript
and it is therefore assumed in what follows. 
Details can be looked elsewhere~\cite{Aad:2009wy,CMSPas2009:top02}.
On the other hand $N_{t\bar{t}~\text{dileptons}}$ can be computed using the quantities found in the previous sections,
%allowing the possibility to measure simultaneously $R$ and $\sigma$.
%$N_{t\bar{t}~\text{dileptons}}$ 
and can be parametrized as follows:

\begin{equation}
N_{t\bar{t}~\text{dileptons}} = (1-\alpha_0)\cdot N_{\text{events}} -N_{\text{single~top}} - N_{\text{other}~t\bar{t}}
\label{eq:nttbardileptons}
\end{equation}

where $\alpha_0$ is the probability that no jets from top decays have been reconstructed and selected in the final sample.
By combining Eq.~\ref{eq:crosssection} and Eq.~\ref{eq:nttbardileptons} it is trivial to parametrize $\alpha_0$ as a function of $\sigma$.
It is therefore possible to use this parametrization as an input to the likelihood function in Eq.~\ref{eq:hfcfitlikelihood},
and to leave both $R$ and $\sigma$ to float freely when fitting the $b$-tagging multiplicity distribution.
The result of this combined fit is shown in Fig.~\ref{fig:randcrosssectionfit} and it is compared with the MC input given for both $R$ and $\sigma$.
The result is shown in a detector-independent way, i.e., factorizing the acceptance and efficiencies in the measurement of $\sigma$.
A good agreement is found between the results of the method and the MC expectations, within the statistical uncertainty of the sample used for this study.

The discussion on the robustness of the combined measurement is similar to the discussions
carried through Secs.~\ref{subsec:jetmisassignment} and \ref{subsec:rmeasurement}.
A more detailed study on the final uncertainty on the measurement of the cross section 
would, however, require full detector simulation and estimation on the acceptance and efficiency
of detecting $t\bar{t}$ dilepton events which is out of the scope of this manuscript.

\section{Conclusions}
\label{sec:conclusion}

The physics of the top quark sector is one of the current topics of great interest to study
at the LHC. The scale of the mass of the top quark is set
at approximately the electroweak symmetry breaking scale, 
and according to several BSM models the preferential coupling may mainly occur through the top quark sector.
In this context, the accurate measurement of the top quark properties is crucial to look for evidence of BSM physics
at the LHC.

In this manuscript, a discussion on 
%how 
the measurement of the heavy flavor content of $t\bar{t}$ decays 
%can shed a light on BSM physics evidence in the top quark sector 
was presented.
A specific proposal was suggested to use a likelihood method to fit the $b$-tagging multiplicity distribution,
based on quantities which can be measured directly from data in order to minimize the reliance on MC simulation.
The method is expected to improve the current measurement of $R$, the ratio of branching fractions of the top quark
$R=B(t\rightarrow Wb)/B(t\rightarrow Wq)$, and it can be extended to measure simultaneously also
the production cross section of $t\bar{t}$ events.

For illustration, samples of $t\bar{t}$ dilepton events together with the main backgrounds were used.
It was shown that a simple kinematical variable, such as the lepton-jet invariant mass, can be used to distinguish signal and background events.
A reliable modeling of the background %mis-assigned lepton-jet event pairs 
can be achieved without the use of MC samples, and can be instead directly derived from data.
By using an iterative method, in which part of the unknown variables are measured in one sub-sample
and another (i.e. $R$) is measured in the complementary sample, the systematic uncertainties can be controlled.
Keeping in mind that a more accurate determination of the systematic uncertainty requires measurements from data
and comparison with full detector simulation which were not used in this manuscript,
the results here shown are, however, expected to hold.
The case was studied for a number of $t\bar{t}$ events which is comparable to an equivalent integrated luminosity 
of $\mathcal{L}=$100~pb$^{-1}$ at $\sqrt{s}$=10~TeV. % the systematic uncertainty is smaller than the statistical uncertainty.
%For the $\mathcal{L}=$100~pb$^{-1}$ considered in this manuscript 
The measurement of $R$ is expected to be mostly statistically dominated with an uncertainty of $\approx $~2\%. 
Using the measurements obtained from other $t\bar{t}$ channels, namely the ``lepton+jets'' channel,
and in different control regions, namely different jet multiplicity bins in the same $t\bar{t}$ channel,
the systematical uncertainties are expected to cancel out, or to be largely reduced.
A measurement of $R$ sensitive to deviations from the SM %BSM Physics or confirm the SM 
seems possible with early data at the LHC.

\medskip

{\large \bf Aknowledgements} 

The authors wish to aknowledge Claudio Campagnari, Tim Christiansen and Jo\~ao Varela
for the useful discussions throughout the preparation of the work presented in this manuscript. 
This work was partially supported by the grant SFRH/BD/21484/2005 from Funda\c{c}\~{a}o Ci\^{e}ncia e Tecnologia 
- Minist\'{e}rio da Ci\^{e}ncia Tecnologia e Ensino Superior - Portugal.

\end{document}